\documentclass[a4paper, 11pt]{article}
\pdfoutput=1 

\usepackage{jcappub} 

\usepackage[T1]{fontenc} 
\usepackage{caption}
\usepackage{subcaption}
\usepackage{bbm}

\title{\textbf{Perturbations in pseudo-Nambu-Goldstone Higgs Inflation}}

\author[a]{Stephon Alexander,}
\author[a]{Humberto Gilmer,}
\author[a]{and Cooper Niu}

\affiliation[a]{Department of Physics, Brown University, Providence, RI, 02912, USA}

\emailAdd{stephon\_alexander@brown.edu, humberto\_gilmer@brown.edu, cooper\_niu@brown.edu}

\abstract{Pseudo-Nambu-Goldstone (pNG) Higgs Inflation is a novel approach to relate the Higgs boson and its interaction with Electroweak gauge bosons with cosmic inflation, with the potential of solving both the fine-tuning issues in the Higgs mass and inflationary potentials. In this work, we present a linear perturbation analysis of the minimal implementation of pNG Higgs inflation using the symmetry coset SU($5$)/SO($5$). Similar to Chromo-natural inflation, this model exhibits a period of instability in the tensor modes that exponentially enhance left-handed gravitational waves. Thus, large Chern-Simons couplings $\beta \gtrsim 6 \times 10^8$ and decay constants $f\gtrsim1\times10^{18}~\mathrm{GeV}$ are required to suppress the tensor-to-scalar ratio $r$ to be compatible with the cosmic microwave background (CMB) measurement. These large couplings also cause an overproduction of the scalar modes, making the minimal construction of pNG Higgs inflation disfavored by CMB observations. However, this tension could potentially be relieved by considering multi-field inflation. The pNG Higgs construction naturally contains multiple scalar fields via the interplay of spontaneously broken global and gauge symmetries. The rich structure enables a broad range of multi-field inflation, and we conclude by briefly discussing this possibility and future work.
}

\begin{document}
\maketitle
\flushbottom
\section{Introduction}
Cosmic inflation \cite{Guth:1980zm, Albrecht:1982wi, Linde:1981mu} is a well-studied possible paradigm for the very early universe. The generic form of inflation consists of a slowly-rolling scalar field, the inflaton, that drives a period of exponentially-fast expansion before the Big Bang. The primordial inhomogeneities caused by quantum fluctuations are then diluted and smoothed during the expansion of spacetime. This solves the horizon, flatness, and monopole problems in cosmology and provides a robust mechanism for the large-scale structure formation.
  
As the only fundamental scalar in the Standard Model (SM), the Higgs boson is a compelling candidate for being the inflaton without introducing exotic physics. However, the Higgs self-interaction measured at the LHC $\lambda \sim 0.13$ \cite{LHCHiggsCrossSectionWorkingGroup:2016ypw} far exceeds the constraint $\lambda \sim 10^{-13}$ \cite{Linde:1983gd}, leading to either inflation of insufficient duration or a strong tension with observations of the matter power spectrum. The idea of ``Higgs as the inflaton'' was made possible by non-minimally coupling the Higgs to gravity~\cite{Bezrukov:2007ep}. The non-minimal paradigm makes a robust prediction of the CMB power spectrum but violates perturbative unitarity below the inflationary energy scale \cite{Burgess:2009ea,Barbon:2009ya}.

Pseudo-Nambu-Goldstone (pNG) Higgs inflation \cite{Alexander:2023flr, Alexander:2022uiz} offers a new way to realize the Higgs-as-inflaton by focusing on a shared disease of both the Higgs and inflation: \textsl{fine-tuning problems}. Successful inflation requires a suitably flat potential such that the potential energy of the inflaton dominates over its kinetic energy for a sufficient period of time. However, scalar fields can receive large radiative corrections that ruin the flatness of the potential. Without the protection of symmetries, retaining a slow-rolling inflationary potential requires fine-tuning. This is known as \textsl{the} $\eta$\textsl{ problem.} Similarly, the hierarchy problem for the Higgs mass originates from the huge discrepancy between the LHC measurement $m_{\rm H} = 125 ~\mathrm{GeV}$ and the large quantum corrections that may naturally, in the absence of protective symmetries, go up to the Planck scale $\Lambda \sim 10^{19}~\mathrm{GeV}$. 

Introducing new symmetries is a common way to protect scalars from quadratically-divergent corrections. Natural inflation \cite{Freese:1990rb} uses an axion with a shift symmetry to maintain naturalness, protecting the potential from large quantum corrections. This original form of natural inflation requires a super-Planckian axion decay constant $f$ to match with CMB observations \cite{Freese:2004un}. Chromo-natural inflation \cite{Adshead:2012kp} relieves this issue by decelerating the axion via friction provided by a Chern-Simons coupling term between the axion and a $\mathrm{SU(2)}$ gauge boson. The Chern-Simons term also sources chiral gravitational waves with rich phenomenology.

Similarly, the Higgs mass can also be stabilized against large quadratic corrections by introducing new symmetries. Notably, the paradigm of the little Higgs treats the Higgs as the pseudo-Nambu-Goldstone boson (pNGB) of some strongly-coupled larger symmetry theory \cite{Schmaltz:2005ky, Perelstein:2005ka}. Through an interplay between spontaneously and explicitly broken symmetries, the SM Higgs arises as an angular excitation of the larger, spontaneously broken symmetry group; the mass parameter arises due to the explicit breaking and is protected from large corrections by the larger symmetry \cite{Perelstein:2005ka, Schmaltz:2005ky}. In addition, the angular excitations, including the Higgs, experience a periodic potential with a shift symmetry, which makes a Higgs-driven natural inflation feasible. In \cite{Alexander:2022uiz} the pNG Higgs arising in the `littlest Higgs' model \cite{Arkani-Hamed:2002ikv} built on an SU(5)/SO(5) coset thus serves as a possible inflaton candidate, due to its nature as a pseudo-Nambu-Goldstone boson experiencing a periodic potential, paralleling the axion in natural inflation. In the spirit of the chromo-natural inflation, a coupling between the Higgs and the Chern-Simons current is postulated to provide extra friction to the rolling of the Higgs on steep potentials.

The original Chromo-natural inflation overproduces gravitational waves due to the instabilities in the tensor modes of curvature perturbations and is thus ruled out by CMB measurements \cite{Adshead:2013nka}. To address the inconsistency with CMB observations, a number of scenarios are proposed, including multiple-axion inflation \cite{Obata:2014loa}, nonlinear perturbations \cite{Papageorgiou:2018rfx}, warm inflation \cite{Mukuno:2024yoa,Yeasmin:2022ncm}, and non-minimal coupling to gravity \cite{Dimastrogiovanni:2023oid, Murata:2024urv}. \cite{Adshead:2016omu} introduces a Higgs field that spontaneously breaks the $\mathrm{SU(2)}$ gauge symmetry. The Goldstone mode of the Higgs enhances the scalar perturbations and consequently decreases the tensor-to-scalar ratio $r$ to be compatible with the CMB observation. Our model naturally presents a Higgs and a broken $\mathrm{SU(2)}$ gauge symmetry in the minimal setup. 

In this paper, we present a full cosmological perturbation analysis of the $\mathrm{SU(5)/SO(5)}$ pNG Higgs inflation \cite{Alexander:2023flr, Alexander:2022uiz}. We find that the minimal pNG Higgs inflation is disfavored by CMB measurements as the overproduction of gravitational waves persists. Unlike the Higgsed CNI, the energy scale of this model is fixed to the electroweak scale and hence is less flexible. However, the architecture of the $\mathrm{SU(5)/SO(5)}$ symmetry coset naturally contains 12 more pNGBs, opening up a whole variety of multi-field inflations without introducing new structure. In what follows, we discuss several feasible scenarios and leave the details for future studies.

\textit{Notation Conventions:} We use the natural units where $\hbar = c = M_{\mathrm{Pl}} = (8\pi G)^{-1/2}~=~1$. Our spacetime metric uses the $(-,+,+,+)$ convention.  For convenience, we work with the dimensionless time variable $x = -k\tau$ with $k$ the Fourier mode wavenumber and $\tau$ the conformal time. The Levi-Civita tensor is $\epsilon^{0123} = 1/\sqrt{-g}$. Throughout the analysis, we use dot for cosmic time derivative $(\dot ~) \equiv (\partial_t)$ and prime for dimensionless time derivative $(')\equiv(\partial_x)$. 

\section{pNG Higgs Inflation} 
We focus on the littlest Higgs inflation proposed in \cite{Alexander:2023flr}, where the Higgs emerges as a complex doublet in an $\mathrm{SU(5)/SO(5)}$ nonlinear sigmas model. The effective action of the littlest Higgs inflation is given by 
\begin{align}
\label{eq:model_action}
    \begin{aligned}
        \mathcal{S} = \int \mathrm{d}^4x \sqrt{-g}\Bigg[\frac{R}{2}-\frac{1}{2}(\partial_{\mu}h)^2 &- V(h) -\frac{1}{4}F_{\mu\nu}^aF_a^{\mu\nu} -\frac{1}{8}\left(\frac{g\beta}{f}\right)^2h^2F^a_{\mu\nu}\Tilde{F}_a^{\mu\nu} \Bigg],
    \end{aligned}
\end{align}
where $h$ is the Higgs with a symmetry-breaking scale $f$ and a dimensionless Chern-Simons coupling $\beta$. The tree-level Higgs potential takes a sinusoidal form
\begin{align}
\label{eq:higgs_potential}
    V(h) = \mu^4\left[c_0+\sin ^2\left(\frac{h}{f}\right)+\sin ^4\left(\frac{h}{f}\right)\right]
\end{align}
with a $\mathcal{O}(10^{-2})$ dimensionless constant
$c_0$, and we choose $c_0 = 0.05$ fiducially. To predict the correct Higgs mass $m_H \approx 125~\mathrm{GeV}$, the strong energy scale $\mu$ is related to the decay constant by the relation
\begin{align}
    m_H^2 = 8(\mu^2/f)^2.
\end{align}\label{eq:higgs-mass}
The $\mathrm{SU}(2)$ gauge field strength and its Hodge dual are defined as
\begin{align}
    F_{\mu \nu}^a=\partial_\mu A_\nu^a-\partial_\nu A_\mu^a + g\epsilon^{abc} A_\mu^b A_\nu^c, \quad\text{ and }\quad \Tilde{F}_{\mu\nu}^a = \frac{1}{2}\epsilon^{\mu\nu\alpha\beta}F_{\alpha\beta}^a,
\end{align}
with $\epsilon^{\mu\nu\alpha\beta}$ the Levi-Civita tensor. 

Since the pNG Higgs inflation generally produces an extremely large number of $e$-folds (see figure~\ref{fig:parameter_space}), the last $60$ $e$-folds of perturbation evolution generically constitute a small fraction of the total evolution. The radial mode of the pNG Higgs is approximately non-dynamical at its VEV. Additionally, since the Higgs is the slow-roll inflaton in our model, the dynamics due to the Higgs mass is negligibly small. In light of these, we work in the Stueckelberg limit \cite{Ruegg:2003ps, Grosse-Knetter:1992tbp}, in which the Higgs mass is taken to be infinitely large. The action for the Higgs Goldstone modes can be written as 
\begin{align}
    \mathcal{S}_\text{Goldstone} = \int \mathrm{d}^4x \sqrt{-g} \left[-g^2 v^2 \mathrm{Tr}\big(A_\mu+\partial_\mu\xi\big)^2 \right],
\end{align}
where $\xi = \xi^aJ_a$ is the Higgs angular degree of freedom, $J^a$ are the generators of $\mathrm{SU(2)}$ in the spinor representation, and the Higgs vacuum expectation value (VEV) is fixed by the electroweak scale  $v_\text{EW} \approx 246 ~\mathrm{GeV}$ \cite{ATLAS:2012yve}. To reiterate, note that the $\mathrm{SU(2)}$ gauge boson here is the Standard Model weak isospin boson.

\subsection{Background Solutions}\label{sec:bacground}
To have an isotropic cosmology, we take the Higgs field to be classical and homogeneous $h=h(t)$. We also choose a classical and rotationally-invariant gauge-field configuration:
\begin{align}
A_0^a=0, \quad A_i^a= \phi(t)\delta_i^a = a\psi(t)\delta_i^a,
\end{align}
The equations of motion for the Higgs and the gauge field are 
\begin{align}
    \ddot h + 3H \dot h+V_{,h} &= -3g\left(\frac{g\beta}{f}\right)^2\psi^2h\left(\dot\psi+H\psi\right), \\ 
    \ddot{\psi}+3 H \dot{\psi}+\left(\dot{H}+2 H^2\right) \psi+2 g^2 \psi^3 + g^2v^2 \psi &= g\left(\frac{g\beta}{f}\right)^2 h\dot{h} \psi^2 .
\end{align}
During the slow-roll inflation, the field equation can be approximated as 
\begin{align}
    \dot{\psi} & =-H \psi-\frac{f^2V_{,h}}{3 g^3 \beta^2 h \psi^2} \label{eq:gauge-bg}, \\
    \left(\frac{g\beta}{f}\right)^2 h\dot{h} & = 2g\psi + \frac{gv^2}{\psi}+ \frac{2H^2}{g\psi}. \label{eq:higgs-bg}
\end{align}
\\
\begin{figure}[h]
    \centering
    \begin{subfigure}[b]{0.48\textwidth}
        \centering
        \includegraphics[width=\textwidth]{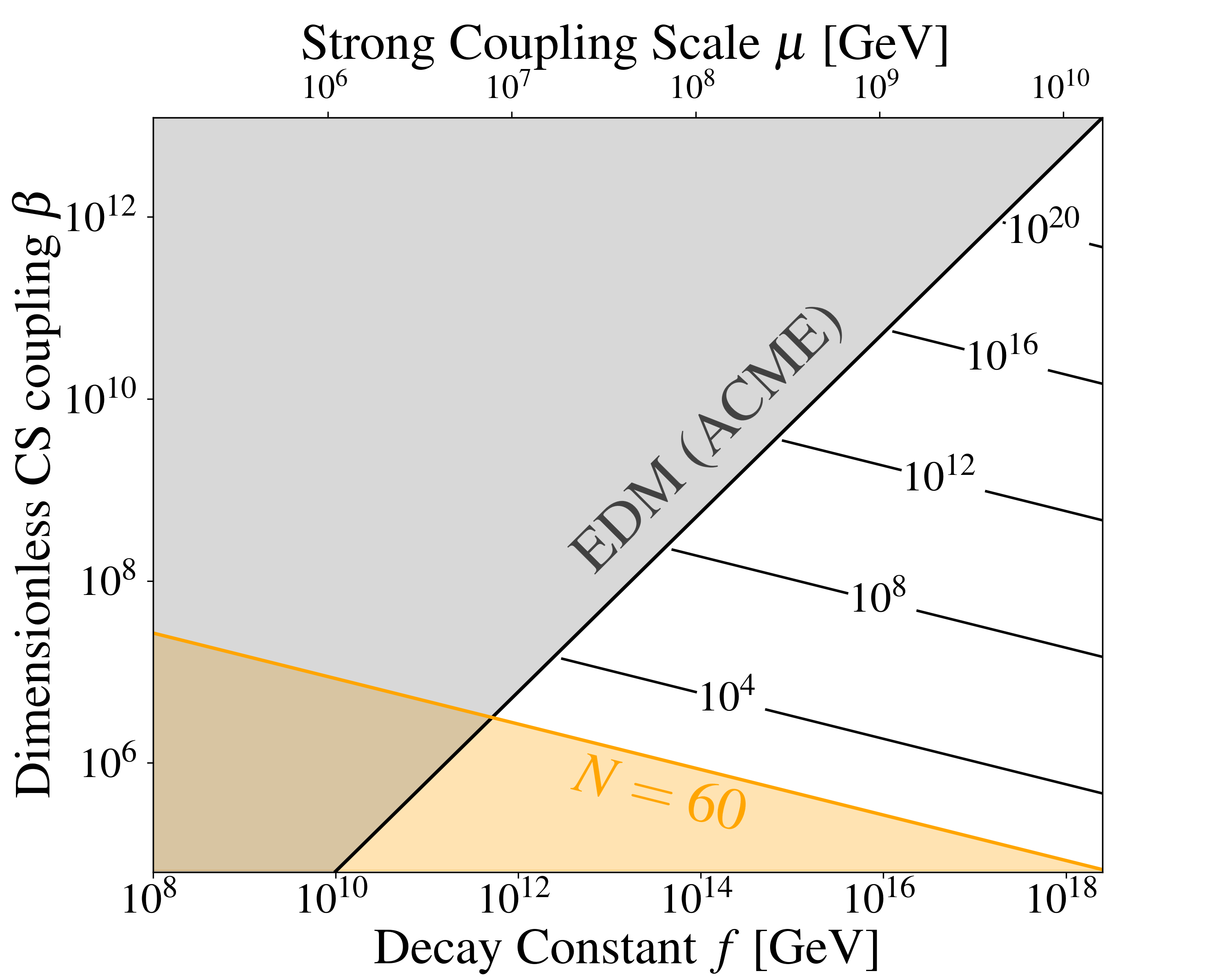}
        \label{fig:mpsi_plot}
    \end{subfigure}
    \hspace{-0.8cm}
    \begin{subfigure}[b]{0.48\textwidth}
        \centering
        \includegraphics[width=\textwidth]{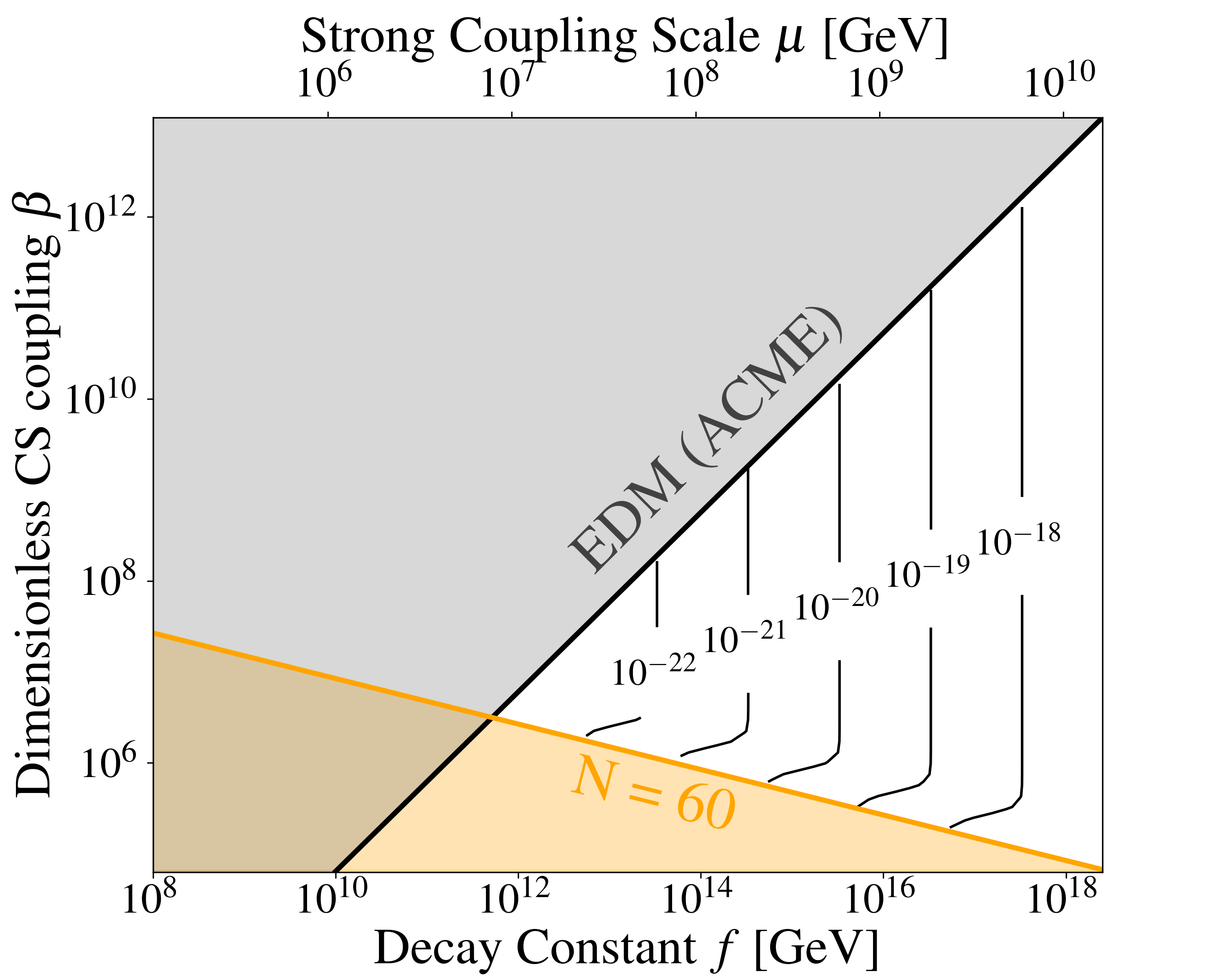}
        \label{fig:mvev_plot}
    \end{subfigure} 
    \begin{subfigure}[b]{0.48\textwidth}
        \centering
        \includegraphics[width=\textwidth]{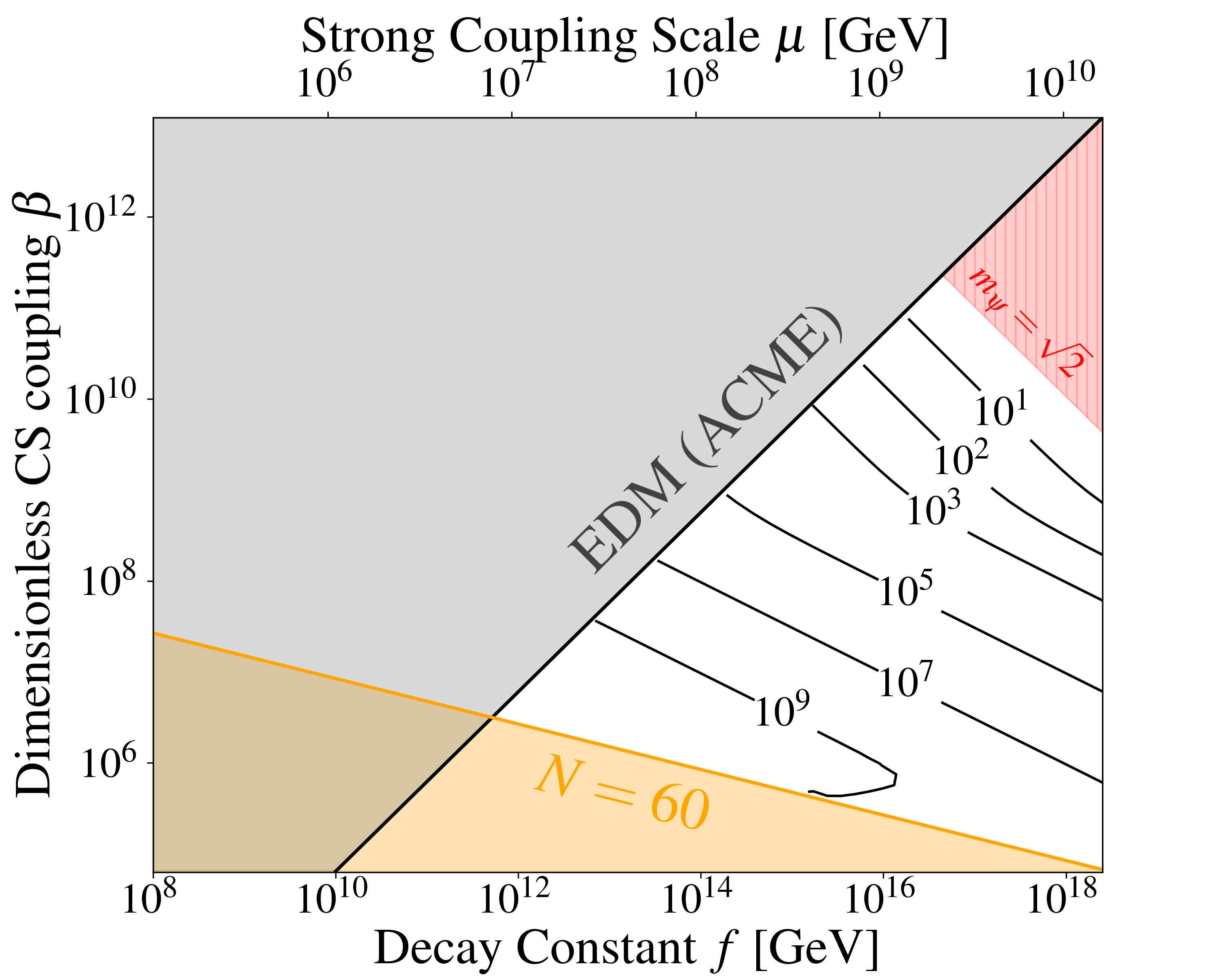}
        \label{fig:mpsi_plot}
    \end{subfigure}
    \hspace{-0.8cm}
    \begin{subfigure}[b]{0.5\textwidth}
        \centering
        \includegraphics[width=\textwidth]{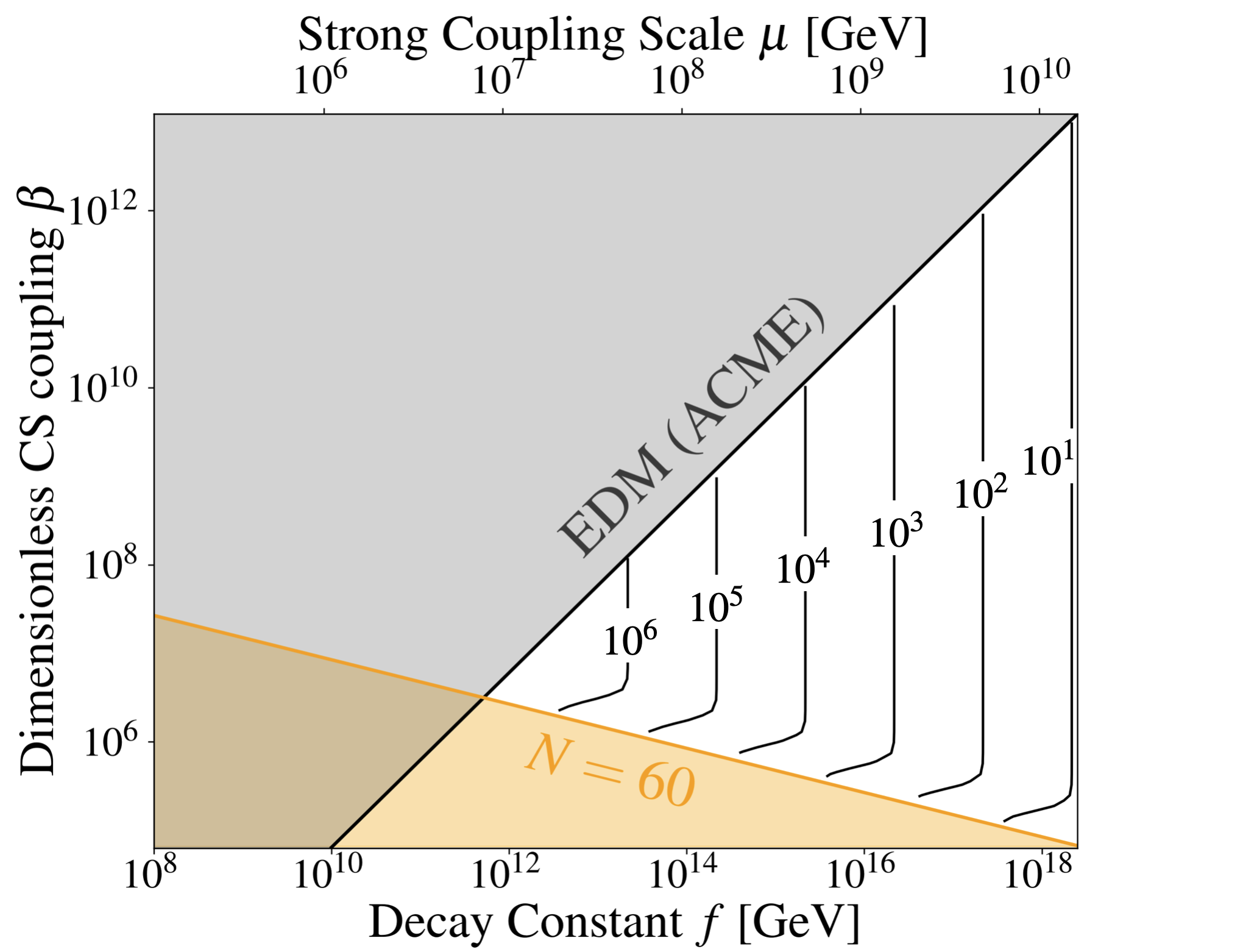}
        \label{fig:mvev_plot}
    \end{subfigure} 
    \caption{The contour plots for the number of $e$-folds (upper left), the Hubble scale (upper right) when the background fields cross the horizon,  $m_\psi$ (lower left), and $m_v$ (lower right). Assuming the decay constant $f < M_\mathrm{Pl}$, the viable parameter space is shown in the white region. The gray region is ruled out by electric-dipole moment (EDM) measurement \cite{Lue:1996pr, Zhang:1993vh}, and the orange region represents the inflation less than $60$ $e$-folds. The red shaded region indicates the region with $m_\psi < \sqrt{2}$, which induces scalar instability. }
    \label{fig:parameter_space}
\end{figure}\\
\noindent

As the gauge field dynamics is negligibly small, we can have
\begin{align}
    \psi \approx \left(-\frac{f^2 V_{,h}}{3 h g^3 H \beta^2}\right)^{1 / 3} \label{eq:gauge-bg2}.
\end{align}
For convenience, we define two dimensionless mass parameters $m_\psi = g\psi/H$ and $m_v = gv/H.$ These two parameters contribute to the effective mass of the gauge field, and we will use them to characterize the instabilities in the scalar and tensor modes for the perturbation analysis. 

As we will elaborate in Sec.~\ref{sec:tensor}, the effective gauge masses $m_\psi$ and $m_v$ are crucial in controlling the tensor instability. To avoid overproducing tensor fluctuations, one must require $m_\psi, m_v \lesssim \mathcal{O}(10)$. We evaluate $m_\psi$ and $m_v$ to understand their dependences on the free parameters $\{f, \mu, \beta\}$. Since the horizon-crossing occurs near the end of the inflation, we expand the potential near its minimum
\begin{align}
    V(h) \approx \mu^4 \left(c_0 + c_2(\theta - \pi)^2\right),
\end{align}
where we define the Higgs angle $\theta \equiv h/f$ and ignore the higher-order contribution from $\sin^4(h/f)$. Hubble scale during the slow-roll inflation is given by
\begin{align}
    H \approx \frac{\mu^2}{\sqrt{3}}\sqrt{c_0 + c_2(\theta-\pi)^2} \approx \mu^2\sqrt{\frac{c_0}{3}}.
\end{align}
For the region of interest, we can safely discard $(\theta-\pi)^2$ because the pNG Higgs is sufficiently close to its minimum. The Hubble scale hence depends only on $\mu$, as shown in figure~\ref{fig:parameter_space}. The effective gauge masses $m_\psi$ and $m_v$ can be approximated as
\begin{align}
    m_\psi \approx  \alpha_0 (\pi - \theta)^{1/3}, \quad m_v = \frac{gv}{\mu^2}\sqrt{\frac{3}{c_0}}
\end{align}\label{eq:m_approx}
with $\alpha_0 = \left(\frac{3}{\pi\beta^2 \mu^4c_0^2}\right)^{1/3}.$ The number of $e$-folds are then given by
\begin{align}
    \Delta N = \int_{\theta_c}^{\pi} d\theta ~ \beta_0 \theta \frac{m_\psi}{2+ 2m_\psi^2 + m_v^2} \approx \frac{3\pi\beta_0}{2\alpha_0^3}\left(2\Tilde{m}_\psi^2 - \Tilde{m}_v^2 \log\left[1 + 2\Tilde{m}_\psi^2/\Tilde{m}_v^2\right]\right)
\end{align}
where $\beta_0 = 2\left(\frac{g\beta}{f}\right)^2$ and $\Tilde{m}_\psi$ and $\Tilde{m}_v$ are effective gauge masses after horizon crossing. For $\Tilde{m}_\psi \sim\Tilde{m}_v$, we determine the order of magnitudes of $\Tilde{m}_\psi$ by dropping the logarithmic term 
\begin{align}
    \Tilde{m}_\psi \sim \sqrt{\frac{40 \alpha_0^3 }{\pi \beta_0}} \approx 77.05\frac{f}{\beta^2 \mu^2}.
\end{align}\label{eq:mpsi-approx}

The requirement that $m_v \lesssim \mathcal{O}(10)$ sets the strong coupling scale to $\mu \gtrsim 10^{10}~\mathrm{GeV}$, which corresponds to $f \gtrsim 10^{18}~\mathrm{GeV}$. In order to have $m_\psi \sim \mathcal{O}(10)$, one needs to
increase the CS coupling $\beta$ to suppress $m_\psi$. However, the stability condition for the scalar perturbations imposes a lower bound $m_\psi > \sqrt{2}$. A healthy theory with minimal instability resides at $\beta \sim 10^9$. At this stage, all the free parameters are confined in a small region by the scalar and tensor stability requirements. This greatly limits the parameter space for pNG Higgs inflation and precludes possibility of reconciling the scalar perturbation power spectrum $A_s$ and the tensor-to-scalar ratio.

\subsection{Quadratic Lagrangian}
To study the dynamics of the linear perturbations, we expand the Lagrangian to quadratic order by considering the following field redefinitions
\begin{align}
   \begin{aligned}
    h(x) &= \Bar{h}(\tau) + \delta h(x) \\
    A_\mu(x) &= \Bar{A}_{\mu}(\tau) + \delta A_\mu(x)  \\
    d s^2 &= -a^2d\tau^2 + a^2\big(\delta_{ij}+\gamma_{ij}+\frac{1}{2}[\gamma^2]_{ij}\big)dx^idx^j 
   \end{aligned}
\end{align}
where the over-bar represents the background fields and $\delta h, \Psi_\mu$, and $\gamma_{ij}$ are the perturbations for the pNG Higgs, the vector boson, and the metric, respectively. The full quadratic Lagrangian is the summation of the Higgs fluctuation,
\begin{align}
    \delta^2\mathcal{L}_h = \frac{a^2}{2}(\partial_\tau\delta h)^2-\frac{a^2}{2}\left(\partial_i\delta h\right)^2 - \frac{a^4}{2}\frac{d^2V}{dh^2}\delta h^2 ,
\end{align}
the quadratic Yang-Mills term, 
\begin{align}
   \begin{aligned}
        \delta^2 \mathcal{L}_{\mathrm{YM}} &=  \operatorname{Tr}\left[\left(\partial_i \Psi_0-i g \phi\left[J_i, \Psi_0\right]\right)^2\right]-4 i g \partial_\tau \phi \operatorname{Tr}\left[\Psi_0\left[\Psi_i, J_i\right]\right] \\
        &\quad +\operatorname{Tr}\left[\partial_\tau \Psi_i \partial_\tau \Psi_i\right]-\operatorname{Tr}\left[\partial_j \Psi_i \partial_j \Psi_i-\partial_i \Psi_j \partial_j \Psi_i\right]+2 g \phi \epsilon_{i j k} \operatorname{Tr}\left[\partial_i \Psi_j \Omega_k\right] \\
        &\quad -g^2 \phi^2 \operatorname{Tr}\left[\left(\Omega_k-\Psi_k\right) \Omega_k\right] -2 \operatorname{Tr}\left[\Psi_0 \partial_\tau\left(\partial_i \Psi_i-i g \phi\left[J_i, \Psi_i\right]\right)\right],
   \end{aligned}
\end{align}
the quadratic Chern-Simons term,
\begin{align}
   \begin{aligned}
        \delta^2 \mathcal{L}_{\mathrm{CS}}= & g \phi^2 \left(\frac{g\beta}{f}\right)^2(\Bar{h}\delta h) \operatorname{Tr}\left[\partial_i \Psi_0 J_i\right]
        -{\left(\frac{g\beta}{f}\right)^2 \Bar{h}}\partial_\tau \Bar{h} \operatorname{Tr}\left[g \phi \Psi_i \Omega_i-\epsilon_{i j k} \Psi_i \partial_j \Psi_k\right]\\
        &+g \phi^2 {\left(\frac{g\beta}{f}\right)^2 \partial_\tau (\Bar{h}\delta h )} \operatorname{Tr}\left[\Psi_i J_i\right]  
        - {\left(\frac{g\beta}{f}\right)^2} \epsilon_{i j k} \partial_\tau \phi {(\Bar{h}\delta h )} \operatorname{Tr}\left[J_i \partial_j \Psi_k\right] \\
        &{+ \frac{3}{2}g\phi^2\left(\frac{g\beta}{f}\right)^2 \delta h^2(\partial_\tau\phi)},
   \end{aligned}
\end{align}
the gauge-metric mixing term,
\begin{align}
    \delta^2 \mathcal{L}_\gamma= & \frac{a^2}{4}\left(\dot{\phi}^2-g^2 \frac{\phi^4}{a^2}\right) \gamma^2  -a^2\left(\frac{\dot{\phi}}{a} \partial_\tau \Psi_{j l}-g \frac{\phi^2}{a^2}\left(2 \epsilon_{i j}^a \partial_{[i} \Psi_{l]}^a+g \phi \Psi_{j l}\right)\right) \gamma_{j l},
\end{align}
and the Goldstone mode contribution
\begin{align}
    \begin{aligned}
             \delta^2 \mathcal{L}_{\xi}=a^4\Big(-\frac{g^2 v^2}{2} g^{\mu \nu}\left(\partial_\mu \xi^a+\Psi_\mu^a\right)\left(\partial_\nu \xi^a+\Psi_\nu^a\right) &+\frac{g^2 v^2 g \psi}{a} \epsilon_{b i c} \xi^b \partial_i \xi^c \\
     &-\frac{g^2 \psi^2 v^2}{4} \gamma^2+g^2 \frac{v^2 \psi}{a} \gamma_{i j} \Psi_{i j}\Big).
    \end{aligned}
\end{align}
Here, we introduce $\Omega_i=i \epsilon_{i j k}\left[J_j, \Psi_k\right] $ and $\gamma^2 = \gamma_{ij} \gamma_{ij} $ to simplify the expressions.

Note that the scalar, vector, and tensor modes in the spatial components of the gauge fluctuation $\Psi_i$ are still mixed together. We can decouple the modes by performing a scalar-vector-tensor (SVT) decomposition 
\begin{align}
\Psi_i=\left(t_i^a+\epsilon_{i k}^a \chi_k+\delta_i^a \delta \phi\right) J_a,
\end{align}
where $J_a$ are the group generators for $\mathrm{SU(2)}$ in the spinor representation. In total, $\Psi_i$ contributes $9$ degrees of freedom, $5$ from the traceless and symmetric tensor $t_i^a$, $3$ from the vector $\chi_k$, and $1$ from the scalar $\delta\phi$. We can eliminate three by gauge-fixing the $\mathrm{SU(2)}$ symmetry. Specifically, we will work in the non-Abelian Coulomb gauge 
\begin{align}
    \bar{D}_i \Psi_i=\partial_i \Psi_i-i g \phi\left[J_i, \Psi_i\right]=0,
\end{align}
following the convention in \cite{Adshead:2013nka}. This choice also provides an additional constraint on the temporal gauge field components 
\begin{align}
\Psi_0^a=-\frac{4 g \partial_\tau \phi \chi_a + a^2g^2v^2\partial_\tau\xi- g \phi^2 \left(\frac{g\beta}{f}\right)^2 \Bar{h}\partial_a \delta h}{-\partial^2+2 g^2 \phi^2 + a^2g^2v^2}.
\end{align}
The non-Abelian Coulomb gauge imposes a further constraint 
\begin{align}
    \partial_i\left(t_i^a+\epsilon_{i j}^a \chi_j+\delta_i^a \delta \phi\right)=2 g \phi \chi^a,
\end{align}
which can been shown to fix the gauge completely \cite{Adshead:2013nka}. The rotational invariance allows us to choose the wavenumber along the $x^3$ direction and define the two helicities of the transverse traceless tensor
\begin{align}
    t^{ \pm}=\frac{1}{\sqrt{2}}\left(\frac{1}{2}\left(t_{11}-t_{22}\right) \pm i t_{12}\right),
\end{align}
two helicities of the transverse vectors, and a scalar
\begin{align}
    v^{ \pm}=\frac{1}{\sqrt{2}}\left(t_{3,1} \pm i t_{3,2}\right), \quad u^{ \pm}=\frac{1}{\sqrt{2}}\left(\chi_1 \pm i \chi_2\right), \quad z = \frac{1}{6}\left(2 t_{33}-t_{11}-t_{22}\right).
\end{align}
 The non-Abelian Coulomb gauge condition can be simplified into
\begin{align}
    -ik(v_\pm \pm iu_\pm) &= 2g\phi u_\pm,  \label{eq:gaueg-fixing1} \\
    -ik(2z+\delta\phi) &= 2g\phi\chi_3 \label{eq:gaueg-fixing2}.
\end{align}
To sum it up, the gauge fluctuation $\Psi_\mu$ contains only six degrees of freedom after gauge-fixing: $\delta\phi, z, v_\pm, u_\pm$, and $t_\pm$. Notice that we have removed $\chi_3$ using Eq.~(\ref{eq:gaueg-fixing2}). 

Lastly, we decompose the Goldstone modes into a scalar mode $\xi^3$ and two vector modes
\begin{align}
    \xi^{\pm} = \frac{1}{\sqrt{2}}\left(\xi^1 \pm i\xi^2\right).
\end{align}
After the SVT decomposition and gauge-fixing, the full quadratic Lagrangian for the pNG Higgs inflation is rewritten as 
\begin{align}
\begin{aligned}
\delta^2 \mathcal{L}= & {\frac{a^2}{2} \left(\partial_\tau \delta h\right)^2-\frac{a^2}{2} \left(\partial_i \delta h\right)^2- \frac{a^4}{2} \frac{d^2 V}{d h^2} \delta h^2} \\
& + a^2g^2v^2\Psi^a_0\partial_\tau \xi^a + \frac{1}{2}a^2g^2v^2\partial_\tau\xi^a\partial_\tau \xi^a - \frac{1}{2}a^2g^2v^2 \partial_i \xi^a \partial_i \xi^a - a^2 g^2 v^2 (t_i^a + \epsilon_{ij}^a \chi^j + \delta_i^a\delta\phi)\partial_i \xi^a \\ 
& + a^2g^2v^2g\psi \epsilon_{bic} \xi^b \partial_i \xi^c +\frac{1}{2} \Psi_0^a\left(-\partial^2+2 g^2 \phi^2 + a^2g^2v^2\right) \Psi_0^a + g\phi \epsilon_{ba}^i\Psi_0^b\partial_i\Psi_0^a\\
& +\Psi_0^a\left(\partial_\tau \partial_i\left(t_i^a+\epsilon_{i j}^a \chi^j+\delta_i^a \delta \phi\right)-2 \partial_\tau\left(g \phi \chi^a\right) + 4 g \partial_\tau \phi \chi^a - g \phi^2 {\left(\frac{g\beta}{f}\right)^2 \Bar{h}}\partial^a \delta h\right) \\
& +\frac{1}{2} \partial_\tau t_i^a \partial_\tau t_i^a+\partial_\tau \chi_i \partial_\tau \chi_i+\frac{3}{2} \partial_\tau \delta \phi \partial_\tau \delta \phi-\frac{1}{2} \partial_j t_i^a \partial_j t_i^a-\partial_j \chi_i \partial_j \chi_i-\frac{3}{2} \partial_j \delta \phi \partial_j \delta \phi \\
& +\frac{1}{2} \partial_i\left(t_i^a+\epsilon_{i j}^a \chi_j+\delta_i^a \delta \phi\right) \partial_k\left(t_k^a+\epsilon_{k j}^a \chi_j+\delta_k^a \delta \phi\right)+g \phi\left(\epsilon_{i j k} \partial_i t_j^a t_k^a-\epsilon_{i j k} \partial_i \chi_j \chi_k+6 \partial_i \chi_i \delta \phi\right) \\
& -2 g^2 \phi^2\left(2 \chi_i \chi_i+9 \delta \phi^2\right)-\frac{1}{2} g \phi {\left(\frac{g\beta}{f}\right)^2 \Bar{h}}\partial_\tau \Bar{h}\left(t_i^a t_i^a-2 \chi_i \chi_i-6 \delta \phi^2\right) \\
& -\frac{a^2g^2v^2}{2}\left(t_i^at_i^a + 2\chi^i\chi^i + 3\delta\phi^2\right)+3g \phi^2 {\left(\frac{g\beta}{f}\right)^2 (\delta h \delta \phi\partial_\tau \Bar h + \Bar{h}\partial_\tau \delta h\delta \phi)}  +2{\left(\frac{g\beta}{f}\right)^2 \Bar{h}}\partial_\tau \phi \delta h \partial_j \chi_j \\
&+ \frac{1}{2}{\left(\frac{g\beta}{f}\right)^2 \Bar{h}}\partial_\tau \Bar{h}\left(\epsilon_{i j k} \partial_i t_j^a t_k^a-2 \partial_i t_{i j} \chi_j-\epsilon_{i j k} \partial_i \chi_k \chi_j-4 \delta \phi \partial_i \chi_i\right)+ {\frac{3}{2}g\phi^2\left(\frac{g\beta}{f}\right)^2 \partial_\tau\phi\delta h^2} \\
& +{\frac{a^2}{8}\left(\left(\partial_\tau \gamma\right)^2-\left(\partial_i \gamma\right)^2+2\left(\dot{\phi}^2-g^2 \frac{\phi^4}{a^2}\right) \gamma^2\right) -a^2\left(\frac{\dot{\phi}}{a} \partial_\tau t_{j l}-g \frac{\phi^2}{a^2}\left(2 \epsilon_{i j}^a \partial_{[i} t_{l]}^a+g \phi t_{j l}\right)\right) \gamma_{j l}.}
\end{aligned}
\end{align}
In the following analysis, we separate the scalar and tensor contributions in the quadratic Lagrangian and study their dynamics respectively. 

\section{Linear Perturbations}\label{sec:perturbation}
In this section, we perform a comprehensive analysis of the pNG Higgs inflation perturbation to linear order. The dimension-6 Chern-Simons operator modifies the behavior of the scalar perturbations significantly compared to the Higgsed CNI, so we will thoroughly detail the dynamics of these scalar perturbations. It is worth mentioning that the field equations of the vector and tensor perturbations are identical to the Higgsed Chromo-natural inflation, so their dynamics are similar in both systems; consequently we only review the tensor mode analysis briefly. 

\subsection{Scalar Modes}\label{sec:scalar}
In Fourier space, the terms that contribute to the scalar fluctuations are given by 
\begin{align}
    \begin{aligned}
        \delta^2 \mathcal{L}= & \frac{a^2}{2} \partial_\tau \delta h \partial_\tau \delta h^*-\frac{a^2}{2}\left(k^2+a^2 \frac{d^2 V}{d h^2}\right) \delta h \delta h^* + \frac{3}{2}g\phi^2\left(\frac{g\beta}{f}\right)^2 \partial_\tau\phi\delta h \delta h^*\\ 
        &-\frac{1}{2}\frac{1}{k^2+2 g^2 \phi^2{+g^2 a^2 v^2}}\left|4 g \partial_\tau \phi \chi_3+ {a^2 g^2 v^2 \partial_\tau \xi}+i k g \phi^2 \left(\frac{g\beta}{f}\right)^2\Bar{h}\delta h\right|^2 \\
        & +3 \partial_\tau z \partial_\tau z^* +\partial_\tau \chi_3 \partial_\tau \chi^*_3+\frac{3}{2} \partial_\tau \delta \phi \partial_\tau \delta \phi^*-3 k^2 z z^*-k^2 \chi_3 \chi^*_3-\frac{3}{2} k^2 \delta \phi \delta \phi^* \\
        &{+\frac{a^2 g^2 v^2}{2} \partial_\tau \xi \partial_\tau \xi^*-\frac{a^2 g^2 v^2}{2} k^2 \xi \xi^*-\frac{a^2 g^2 v^2}{2}(i k(2 z+\delta \phi) \xi^* - i k(2 z^*+\delta \phi^*) \xi)} \\
        &{-\frac{g^2 a^2 v^2}{2}\left(6 zz^*+2 \chi_3 \chi_3^*+3 \delta \phi \delta \phi^*\right)} -3 i k g \phi\left(\chi_3 {\delta \phi}^*-\delta \phi {\chi}^*_3\right) \\
        & -9 g^2 \phi^2 \delta \phi {\delta \phi}^* -\frac{1}{2} g \phi {\left(\frac{g\beta}{f}\right)^2 \Bar{h}}\partial_\tau \bar{h}\left(6 z z^*-2 \chi_3 {\chi}^*_3-6 \delta \phi {\delta \phi}^*\right)\\
        &+i k {\left(\frac{g\beta}{f}\right)^2\bar{h}} \partial_\tau \bar{h}\left(\left(z {\chi}^*_3-{z}^* \chi_3\right)+\left({\delta \phi}^* \chi_3-\delta \phi {\chi}^*_3\right)\right) \\
        & +\frac{3}{2} g \phi^2 {\left(\frac{g\beta}{f}\right)^2\left(\Bar{h}\partial_\tau \delta h^* \delta \phi+ \Bar{h}\partial_\tau \delta h {\delta \phi}^* + \partial_\tau\Bar{h} \delta h^* \delta \phi + \partial_\tau\Bar{h} \delta h\delta \phi^*\right)} \\
        &-\left(\frac{i k^3}{k^2+2 g^2 \phi^2}\right) {\left(\frac{g\beta}{f}\right)^2 \Bar{h}}\partial_\tau \phi\left(\delta h^* \chi_3-\delta h {\chi}^*_3\right), 
    \end{aligned}
\end{align}
where $^*$ indicates complex conjugation. We diagonalize the kinetic term and eliminate the $\chi_3$ degree of freedom by choosing 
\begin{align}
    \delta \hat{\phi}  =\sqrt{2}\left(\frac{\delta \phi}{2}+z\right) ,\quad
\hat{z}  =\sqrt{2}(z-\delta \phi), \quad \chi_3 = -ik\frac{\sqrt{2}\delta\hat{\phi}}{2g\phi}.
\end{align}
For the convenience of numerical integration, we canonically normalize the perturbations by defining
\begin{align}
    \h = a\delta h, \quad \hat{\varphi} = \delta \hat{\phi}\left(\frac{x^2}{m_\psi^2}+2\right)^{1 / 2} ,\quad \hat{X} = ik\frac{m_v}{x}\sqrt{\frac{2m_\psi^2 +x^2}{m_v^2 + 2m_\psi^2 + x^2}} \xi.
\end{align}
During the slow-roll inflation, the dynamics of the background fields and Hubble parameter are slow-roll suppressed such that we can take the scale factor as 
\begin{align}
    a \approx - \frac{1}{H\tau}. 
\end{align}
In our analysis, we also use a useful quantity 
\begin{align}
    \mathbf{\Lambda} \equiv \left(\frac{g\beta}{f}\right)^2\psi.
\end{align}
As the complete details of the scalar mode action are not crucially informative, we note them in Appendix~\ref{sec:scalareom} for curious readers. Defining the scalar perturbations into a vector form $\Delta = (\h, \hat{\varphi}, \hat{z},\hat{X})$, the equations of motion can be represented in a matrix form 
\begin{align}
    \mathbb{I}\Delta'' + \mathbb{M} \Delta' + \mathbb{N}\Delta = 0
\end{align}
where $\mathbb{I}$ is the identity matrix, $\mathbb{M}$ and $\mathbb{N}$ are coefficients for $\Delta$ and its first derivative. The complications of coefficients and the inter-connecting modes make it unlikely to solve analytically, so we solve the system numerically. 
\begin{figure}[h]
    \centering
    \includegraphics[width=1.0\linewidth]{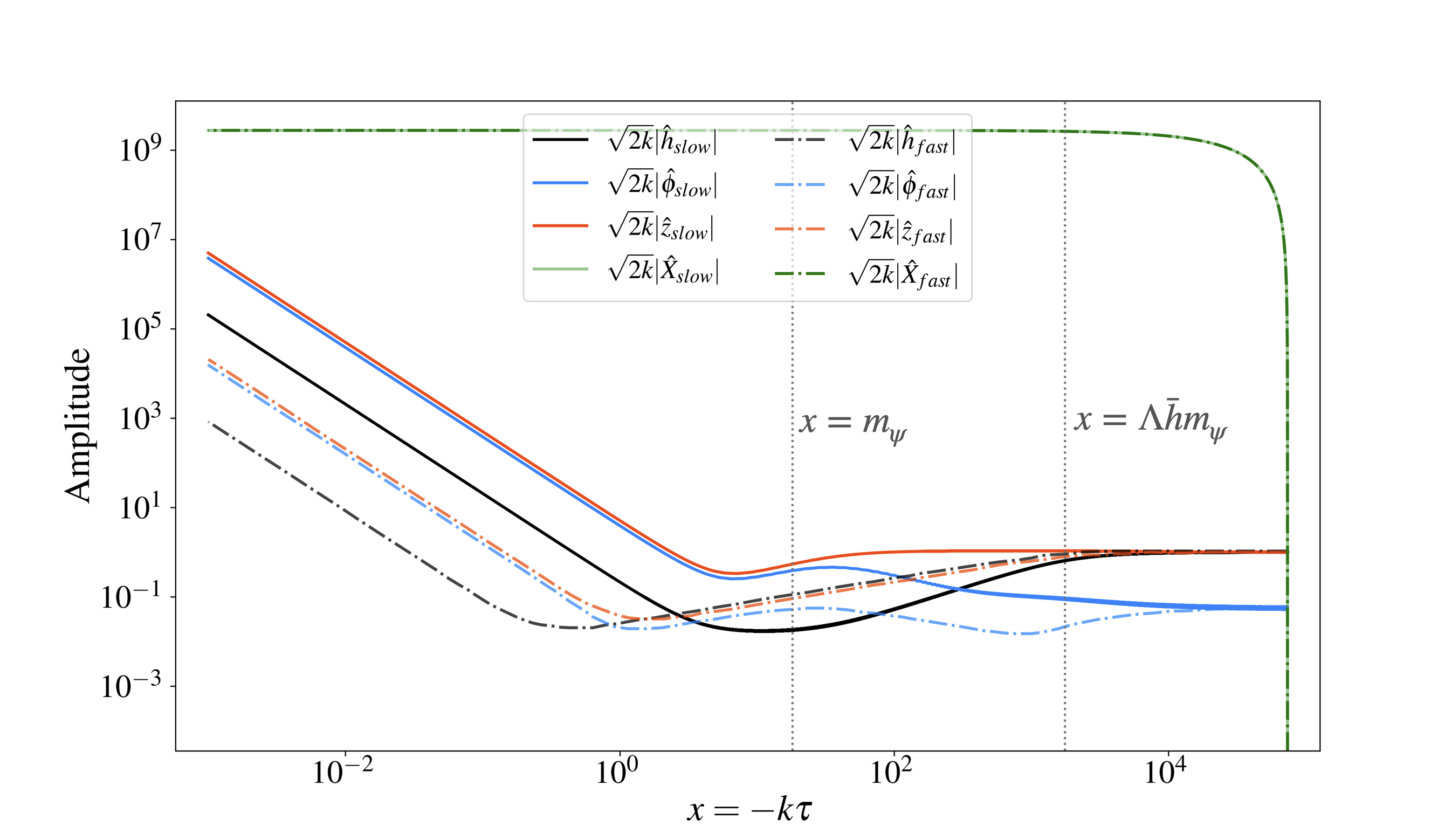}
        \caption{Evolution of scalar perturbations: the little Higgs perturbation $\sqrt{2k}|\h|$ (black), the gauge field perturbations $\sqrt{2k}|\hat{z}|$ (blue), $\sqrt{2k}|\hat{\varphi}|$ (orange), and the Goldstone mode $\sqrt{2k}|\hat{X}|$ (green). The solid lines represent the slow modes, and the dashed lines represent the fast modes. We assume the background freezes at their horizon-crossing values. The parameter choice of this plot is $f=1.8\times 10^{18}~\mathrm{GeV}, \beta=6\times 10^8, \mu \approx 1.1 \times 10^{10}~\mathrm{GeV}$. This specific choice yields $m_\psi \approx 18.2, \psi \approx 1.55 \times 10^{-16}, \bar{h} \approx 2.35, m_v \approx 12.05$, and $H \approx 5.44\times 10^{-18}$.}
    \label{fig:scalar_dynamics}
\end{figure}

We set the initial conditions using the Wentzel-Kramers-Brillouin (WKB) method. In the asymptotic past $(x \to \infty)$, we expand the equations of motion to the leading order of $\frac{\mathbf{\Lambda}}{x}$
\begin{align}
\begin{aligned}
    \h'' + \h + \frac{\sqrt{2}\mathbf{\Lambda}m_\psi \Bar{h}}{x}\hat{z}' - \frac{\sqrt{2}\mathbf{\Lambda}\Bar{h}}{x}\hat{\varphi} &= 0, \\
    \hat{\varphi}'' + \hat{\varphi} - \frac{\sqrt{2}\mathbf{\Lambda}\Bar{h}}{x}\h &= 0, \\
    \hat{z}''+ \hat{z} - \frac{\sqrt{2} \mathbf{\Lambda} m_\psi \bar{h}}{x}\h' &=0, \\
    \hat{X}'' + \hat{X} &= 0.
\end{aligned}
\end{align}
For $x \gg \mathbf{\Lambda}\bar{h} m_\psi$, the four scalar modes are decoupled and freely oscillates. We characterize the asymptotic solution by defining 
\begin{align}
    \h(x)=A e^{i S(x)}, \quad\hat{\varphi}(x)=B e^{i S(x)}, \quad\hat{z}(x)=C e^{i S(x)},\quad \hat{X}(x) = De^{i S(x)}
\end{align}
where we approximate the amplitudes $A, B, C$, and $D$ to be time-independent. The field equations then read
\begin{align}
\begin{aligned}
    \left(1 - (S')^2\right)A + \frac{i\sqrt{2}\mathbf{\Lambda}m_\psi \Bar{h} S'}{x}C -\frac{\sqrt{2}\mathbf{\Lambda}\Bar{h}}{x}B &= 0, \\
    \left(1 - (S')^2\right)B - \frac{\sqrt{2}\mathbf{\Lambda}\Bar{h}}{x}A &= 0, \\
    \left(1 - (S')^2\right)C - \frac{i\sqrt{2} \mathbf{\Lambda} m_\psi \bar{h}S'}{x}A &=0, \\
    \left(1 - (S')^2\right)D &=0.
\end{aligned}
\end{align}
For large $x$, we expect the frequency $S' \approx 1$ since the deviations are suppressed by $x$ or higher powers of $x$. This allows us to disregard $S''$ as it would be suppressed by $x^2$ or higher. We then solve for the eigenfrequencies of the system by letting the determinant of the differential equations to vanish,
\begin{align}
    S' = \pm 1,\pm 1, \pm \left(1+\frac{3}{2} \mathbf{\Lambda}^2\bar{h}^2 \frac{m_\psi^2}{x^2} \pm \frac{\mathbf{\Lambda}\bar{h}}{x} \sqrt{\frac{9}{4} \frac{\mathbf{\Lambda}^2\bar{h}^2}{x^2} m_\psi^4+2\left(1+m_\psi^2\right)}\right)^{1 / 2}.
\end{align}
We discard the $S' = \pm 1$ solution as it requires the Higgs fluctuation amplitude $A$ to be zero. We also focus on the positive frequencies. For the remaining two solutions, we call the one with a plus (minus) sign the fast (slow) mode. Both eigenfrequencies are imaginary if
\begin{align}
    x^2 < \mathbf{\Lambda}^2\Bar{h}^2(2-m_\psi^2),
\end{align}
which causes an exponential production of the scalar perturbations. To avoid this, we impose a stability condition: $m_\psi > 2$. The initial amplitudes of the gauge fluctuations $\hat{\varphi}$ and $\hat{z}$ can be derived
\begin{align}
    \begin{aligned}
        B &= \frac{\sqrt{2}\mathbf{\Lambda}\Bar{h}}{x(1-{S'}^2)} A ,\quad\quad C &=\frac{i\sqrt{2}\mathbf{\Lambda}\Bar{h}m_\psi}{x(1-{S'}^2)} A.
    \end{aligned}
\end{align}
In the limit $x\to\infty$, the eigenfrequencies can be expanded into
\begin{align}
    S' = 1 - \frac{\mathbf{\Lambda}\Bar{h}\sqrt{m_\psi^2+1}}{\sqrt{2}x},
\end{align}
Integrating both sides, we get
\begin{align}
    S = x - x_\infty + \mathbf{\Lambda}\Bar{h}\sqrt{\frac{m_\psi^2+1}{2}} \ln\left(\frac{x_\infty}{x}\right).
\end{align}
For $x =-k\tau \to \infty$, we impose an adiabatic normalization to the Higgs mode by setting $A = 1/\sqrt{2k}$. The initial conditions can thus be written as
\begin{align}
\begin{aligned}
\h_\infty &=\frac{1}{\sqrt{2k}} e^{i\left(x-x_\infty\right)}\left(\frac{x_\infty}{x}\right)^{i \mathbf{\Lambda} \Bar{h}\sqrt{(1+m_\psi^2)/2} }, \\
\hat{\varphi}_\infty &=\frac{1}{\sqrt{2k}}\frac{1}{\sqrt{1+m_\psi^2}} e^{i\left(x-x_\infty\right)}\left(\frac{x_\infty}{x}\right)^{i \mathbf{\Lambda} \Bar{h}\sqrt{(1+m_\psi^2)/2} }, \\
\hat{z}_\infty &=\frac{1}{\sqrt{2k}}\frac{i m_\psi}{\sqrt{1+m_\psi^2}} e^{i\left(x-x_\infty\right)}\left(\frac{x_\infty}{x}\right)^{i \mathbf{\Lambda} \Bar{h}\sqrt{(1+m_\psi^2)/2} }, \\
\hat{X}_\infty &= 0.
\end{aligned}
\label{eq:wkb_initial}
\end{align}
In our numerical evaluation, the initial time $x_\infty$ is chosen to be $2$ orders of magnitude greater than $\Lambda \bar{h} m_\psi$ to ensure the validity of the WKB approximation. 

The evolution of scalar perturbations is presented in figure~\ref{fig:scalar_dynamics}. In the asymptotic past region $x \gg \Lambda \bar{h}m_\psi$, we get free oscillating solutions because the modes are decoupled by large $x$. We notice that the fast mode decays significantly within the horizon so that its to the overall perturbations are negligible comparing the slow mode. It is then justified to ignore the fast mode entirely in our following analysis. 

The curvature perturbation is given by
 \begin{align}
    \mathcal{R} \approx \frac{\delta h}{\dot{\bar{h}}} H.
\end{align}
Recall that the late-time comoving Higgs perturbation evolves as $\sqrt{2k}|\h| \sim \frac{h_0}{x^2}$, and thus the physical Higgs perturbation evolves as
\begin{align}
    \delta h \sim \frac{1}{\sqrt{2k^3}}\frac{Hh_0}{x},
\end{align}
where $h_0$ is obtained numerically. 
From above, the curvature perturbation becomes
\begin{align}
    \mathcal{R} \approx \frac{1}{\sqrt{2k^3}} \left(\frac{g\beta}{f}\right)^2\frac{\bar{h}m_\psi H h_0}{x(2m_\psi^2+m_v^2+2)}.
\end{align}\label{eq:scalar-curvature}
We define the dimensionless power spectrum of curvature perturbation as follows
\begin{align}
\left\langle\mathcal{R}_{\mathbf{k}} \mathcal{R}_{\mathbf{k}^{\prime}}\right\rangle=(2 \pi)^3 \delta^3\left(\mathbf{k}+\mathbf{k}^{\prime}\right) \frac{2 \pi^2}{k^3} \Delta_{\mathcal{R}}^2(k)
\end{align}
The Higgs perturbation $\h$ exits the horizon around $x \sim m_\psi$ and grows as $a^2$ outside the horizon. Since the normalized $\h$ is defined as $\h = a \delta h$, the physical Higgs fluctuation $\delta h$ thus grows as the scale factor $a$, amplifying the scalar power spectrum even outside the horizon. This overproduces the scalar modes, driving up the scalar power spectrum amplitude $A_s$, thus creating a tension with the CMB anisotropy measurements. To maintain a small tensor-to-scalar ratio $r<0.028$, the scalar power spectrum amplitude in this model is at least three orders of magnitude larger than the Planck measurement, $A_s \approx 2.1 \times 10^{-9}$. The parameter scan is shown in figure~\ref{fig:As}. Although it is possible to lower $A_s$ by decreasing the Chern-Simons coupling $\beta$ and decay constant $f$, small $f$ and $\beta$ tend to amplify the tensor modes by enhancing their instabilities. 

\subsection{Tensor Modes}\label{sec:tensor}
The relevant tensor degrees of freedom are the spin-$2$ fluctuations of the gauge field and the gravitational wave 
\begin{align}
\begin{aligned}
\gamma^{ \pm} & =\frac{1}{\sqrt{2}}\left(\frac{1}{2}\left(\gamma_{11}-\gamma_{22}\right) \pm i \gamma_{12}\right), \quad t^{ \pm} =\frac{1}{\sqrt{2}}\left(\frac{1}{2}\left(t_{11}-t_{22}\right) \pm i t_{12}\right) 
\end{aligned}
\end{align}
Canonical normalization requires the following redefinitions
\begin{align}
\hat{\gamma}^{ \pm}=\frac{a \gamma^{ \pm}}{\sqrt{2}}, \quad \hat{t}^{ \pm}=\sqrt{2} t^{ \pm}.
\end{align}
The effective action for tensor modes is thus
\begin{align}
\begin{aligned}
\mathcal{S}_T= \frac{1}{2} &\int \frac{d^3 k}{(2 \pi)^3} d \tau  {\left[\partial_\tau \hat{\gamma}_k^{ \pm} \partial_\tau {\hat{\gamma}}_k^{*\pm}-\left(k^2-\frac{1}{a} \frac{\partial^2 a}{\partial \tau^2}-2 \dot{\phi}^2+2 g^2 \frac{\phi^4}{a^2} + 2g^2a^2v^2\psi^2\right) \hat{\gamma}_k^{ \pm} {\hat{\gamma}}_k^{*\pm}\right.} \\
& + \partial_\tau \hat{t}_k^{ \pm} \partial_\tau {\hat{t}}_k^{*\pm}-\left(k^2+g \phi {\left(\frac{g\beta}{f}\right)^2 \bar{h}\partial_\tau \Bar{h}} + 2g^2a^2v^2\right) \hat{t}_k^{ \pm} {\hat{t}}_k^{*\pm} \pm k\left({\left(\frac{g\beta}{f}\right)^2 \Bar{h} \partial_\tau \Bar{h}}+2 g \phi\right) \hat{t}_k^{ \pm} {\hat{t}}_k^{*\pm} \\
& -2 \dot{\phi}\left(\partial_\tau \hat{t}_k^{ \pm} {\hat{\gamma}}_k^{*\pm}+\partial_\tau {\hat{t}}_k^{*\pm} \hat{\gamma}_k^{ \pm}\right) \mp 2 k g \frac{\phi^2}{a}\left(\hat{t}_k^{\mp} {\hat{\gamma}}_k^{*\pm}+{\hat{t}}_k^{*\pm} \hat{\gamma}_k^{ \pm}\right) \\
& \left.+2 g^2 \frac{\phi^3}{a}\left(\hat{t}_k^{ \pm} {\hat{\gamma}}_k^{*\pm}+{\hat{t}}_k^{*\pm} \hat{\gamma}_k^{\pm}\right) + 2g^2a^2v^2 \psi(\hat{t}^{*\pm} \hat{\gamma}^{ \pm}+\hat{t}^{ \pm} \hat{\gamma}^{*\pm})\right]
\end{aligned}
\end{align}
Working in the slow-roll region, we use the background fields as in Eqs.~(\ref{eq:gauge-bg}) and  (\ref{eq:higgs-bg}) to simplify the action. The equations of motion are,
\begin{align}
\hat{\gamma}_k^{ \pm \prime \prime}+\left(1-\frac{2}{x^2}-\frac{2\psi^2}{x^2}\left(1-m_\psi^2-m_v^2\right) \right) \hat{\gamma}_k^{ \pm} - 2 \frac{\psi}{x} t_k^{ \pm \prime} - 2 m_\psi\left(m_\psi \mp x\right) \frac{\psi}{x^2} t_k^{ \pm} = 0  \label{eq:gw_eom}.
\end{align}
and 
\begin{align}
\begin{aligned}
t_k^{ \pm \prime \prime} &+\left(1 + \frac{2}{x^2} + \frac{2m_\psi^2}{x^2} + \frac{m_v^2}{x^2}\right) t_k^{ \pm} \mp\left(4 m_\psi + \frac{2}{m_\psi}\right) \frac{1}{x} t_k^{ \pm} \\
 & + \frac{2\psi}{x}\hat{\gamma}_k^{ \pm\prime} - \frac{2\psi}{x^2}(1 \mp x m_\psi + m_\psi^2+m_v^2)\hat{\gamma}_k^{ \pm} = 0 \label{eq:spin2_eom}.
\end{aligned}
\end{align}
We remind the reader that these equations are identical to the tensor mode equations in the Higgsed CNI model. At the effective theory level, the pNG Higgs inflation is very similar to the Higgsed CNI model except for the replacement of the dimension-5 operator $\phi F\Tilde{F}$ with the dimension-6 operator $h^2 F\Tilde{F}$. This replacement yields new terms for scalar dynamics via the chain rule. However, the tensor modes (and vector modes) are unaffected with the help of the background fields in eqs.~(\ref{eq:gauge-bg}) and  (\ref{eq:higgs-bg}).

\begin{figure}[h]
    \centering
    \includegraphics[width=0.75\linewidth]{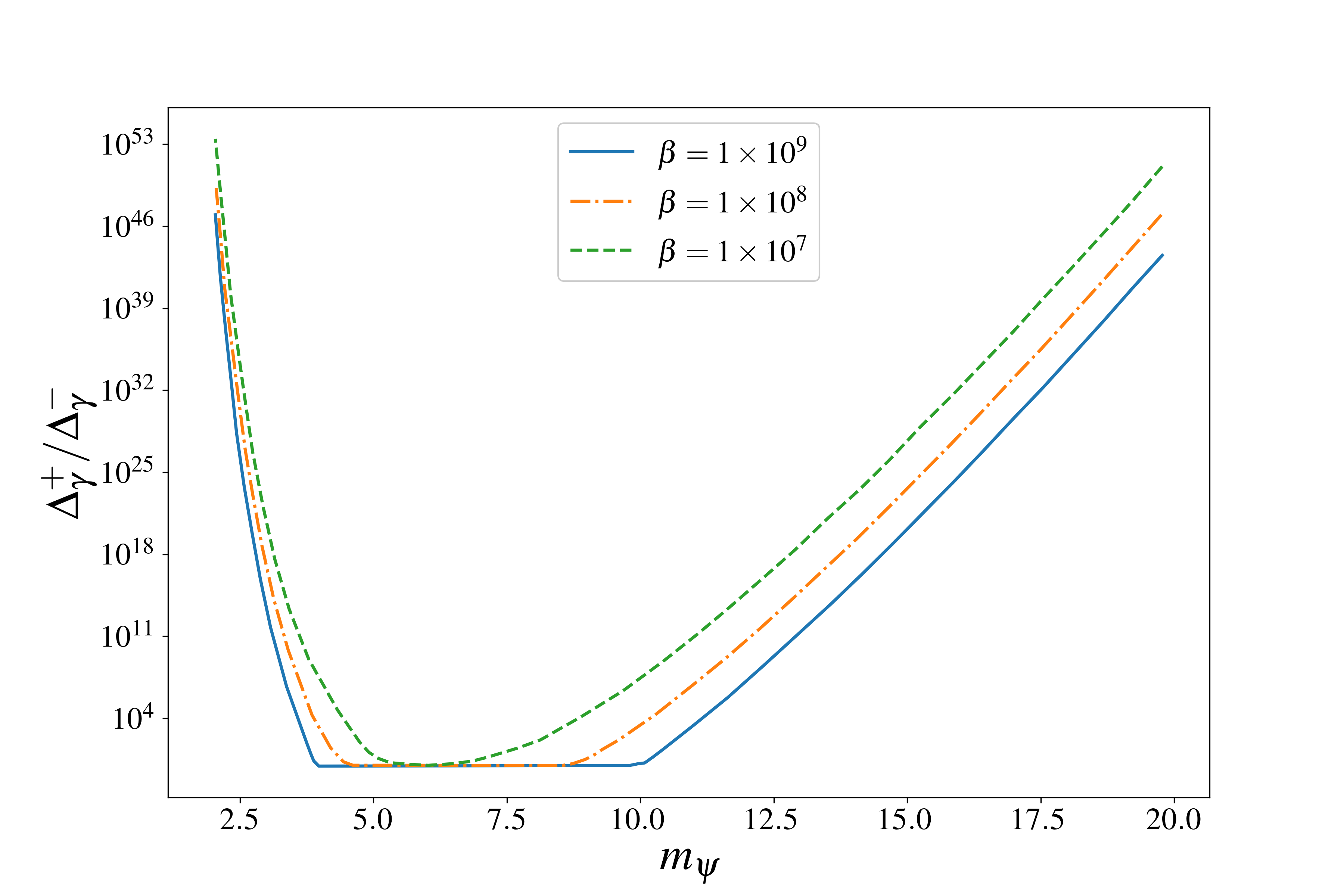}
    \caption{The ratio between left-handed and right handed gravitational wave amplitudes as a function of $m_\psi$. We choose $f = 2\times 10^{18}~\mathrm{GeV}$ right below the Planck scale.}
    \label{fig:tensor_ratio}
\end{figure}

Again, the pNG Higgs inflation prefers longer inflation with a large number of $e$-folds. The regions of interest for perturbation analysis are a tiny fraction of the $e$-folds before the end of the inflation. This allows us to treat the background fields as constant. Moreover, since the pNG Higgs is near the bottom of it potential, the background gauge field takes on values $\psi \ll 1$. The two tensor perturbations are thus decoupled, yielding 
\begin{align}
    t_k^{ \pm \prime \prime}+\left(1+\frac{1}{x^2}(2+2m_\psi^2+m_v^2) \mp \frac{1}{x}\left(4m_\psi+\frac{2}{m_\psi}\right)\right) t_k^{ \pm} & \approx 0, \\
    \hat{\gamma}_k^{ \pm \prime \prime} + \left(1-\frac{2}{x^2}\right)\hat{\gamma}_k^\pm & \approx 0.
\end{align}
 We perform a WKB analysis similar to the scalar modes and find a period of tachyonic instability for the left-handed gauge fluctuation $t_k^+$
 \begin{align}
     2m_\psi + \frac{1}{m_\psi} - \sqrt{2+2m_\psi^2-m_v^2 + \frac{1}{m_\psi^2}} < x < 2m_\psi + \frac{1}{m_\psi} + \sqrt{2+2m_\psi^2-m_v^2 + \frac{1}{m_\psi^2}} \label{eq:instability},
 \end{align}
during which the mode grows exponentially. The $m_\psi$ and $1/m_\psi$ dependencies in Eq.~(\ref{eq:instability}) suggest that the period of instability is significantly elongated for both small and large values of $m_\psi$. This effect is explicit in figure~\ref{fig:tensor_ratio}.

Refs.~\cite{Adshead:2013nka, Adshead:2016omu} have shown that system is analytically solvable, and the late-time solutions for the left-handed and right-handed gravitational waves are 
\begin{align}
    \gamma^+(x) &= \frac{Hx}{\sqrt{k^3}}\left(1+\frac{i}{x}\right)e^{ix} + 2\sqrt{2}\frac{H}{k}B_k\psi (I_1 + m_\psi I_2 - (m_\psi^2 + m_v^2)I_3) \\
    \gamma^-(x) &= \frac{Hx}{\sqrt{k^3}}\left(1+\frac{i}{x}\right)e^{ix} 
\end{align}
The coefficients are
\begin{align}
\begin{aligned}
I_1 &= \frac{\left(m^2-2 i m m_t+2 m-2 m_t^2\right) \sec (\pi \beta) \sinh (-i \pi \alpha) \Gamma(\alpha)}{2 m(m+2)} \\
& ~~~~-\frac{\pi^2\left(m^2+2 i m m_t+2 m-2 m_t^2\right) \sec (\pi \beta) \operatorname{csch}(-i \pi \alpha)}{2 m(m+2) \Gamma(\alpha+1) \Gamma\left(-\alpha-\beta+\frac{1}{2}\right) \Gamma\left(-\alpha+\beta+\frac{1}{2}\right)}, \\
I_2 &=  \frac{\pi \sec (\pi \beta) \Gamma(-\alpha)}{2 \Gamma\left(-\alpha-\beta+\frac{1}{2}\right) \Gamma\left(-\alpha+\beta+\frac{1}{2}\right)}-\frac{\pi \sec (\pi \beta) \Gamma(1-\alpha)}{m \Gamma\left(-\alpha-\beta+\frac{1}{2}\right) \Gamma\left(-\alpha+\beta+\frac{1}{2}\right)} \\
& ~~~~+\frac{\pi m \sec (\pi \beta)-i \pi m_t \sec (\pi \beta)}{2 m \Gamma(1-\alpha)}, \\
I_3 &=  \frac{\pi^2\left(m+i m_t\right) \sec (\pi \beta) \operatorname{csch}(-i \pi \alpha)}{m(m+2) \Gamma(i \alpha) \Gamma\left(-\alpha-\beta+\frac{1}{2}\right) \Gamma\left(-\alpha+\beta+\frac{1}{2}\right)}+\frac{\pi\left(m_t+i m\right) \sec (\pi \beta)}{m(m+2) \Gamma(-i \alpha)} \\
B_k & =\frac{1}{\sqrt{2 k}} \frac{\Gamma\left(-\alpha+\beta+\frac{1}{2}\right)}{\Gamma\left(\alpha+\beta+\frac{1}{2}\right)} 2^\alpha i^{\beta+1}(-i)^{\alpha-\beta},
\end{aligned}
\end{align}
where we introduce $m =2+2m_\psi^2+m_v^2=1/4-\beta^2$ and $m_t =2\left(2 m_\psi+1/m_\psi\right)=-2 i \alpha$. The power spectra for the metric fluctuations are defined as 
\begin{align}
    \left\langle\gamma_{\mathbf{k}}^{ \pm} \gamma_{\mathbf{k}^{\prime}}^{ \pm}\right\rangle=(2 \pi)^3 \delta^3\left(\mathbf{k}+\mathbf{k}^{\prime}\right) \frac{2 \pi^2}{k^3} \Delta_{\gamma^{ \pm}}^2(k)
\end{align}
and the total power spectra is given by $\Delta_\gamma^2(k)=2 \Delta_{\gamma^{+}}^2(k)+2 \Delta_{\gamma^{-}}^2(k)$. 

Figure~\ref{fig:tensor_ratio} shows the ratio of the left-handed and right-handed gravitational wave amplitudes as a function of $m_\psi$. Since the instability only occurs in the left-handed modes, the ratio directly reflects that polarization of the gravitational wave. To avoid the tachyonic production of the gravitational wave, a small $m_\psi$ is preferred. Figure~\ref{fig:parameter_space} shows that $m_\psi$ decreases as the CS coupling $\beta$ increases. This trend agrees with figure~\ref{fig:tensor_ratio} as a larger $\beta$ reduces the gravitational wave production. 

\section{Observational Constraints} \label{sec:cmb}
With the thorough studies of scalar and tensor perturbations, we now map the perturbations of the pNG Higgs inflation to the CMB power spectrum and compare them with Planck collaboration 2018 measurements \cite{Planck:2018jri, Planck:2018vyg}. The constraints on inflation are characterized by three parameters in the primordial power spectrum
\begin{align}
\begin{aligned}
    \ln(10^{10}A_s) &= 3.04 \pm 0.014, \\
    n_s &< 0.9649 \pm 0.0042, \\
    r &< 0.028.
\end{aligned} 
\end{align}
Following the convention of Planck collaboration, we evaluate the tensor-to-scalar ratio, $r$, at $k = 0.002~\mathrm{Mpc}^{-1}$ and the spectral tilt $n_s$ and the scalar amplitude $A_s$ at $k = 0.05~\mathrm{Mpc}^{-1}$.

\begin{figure}[h]
    \centering
    \includegraphics[width=0.7\linewidth]{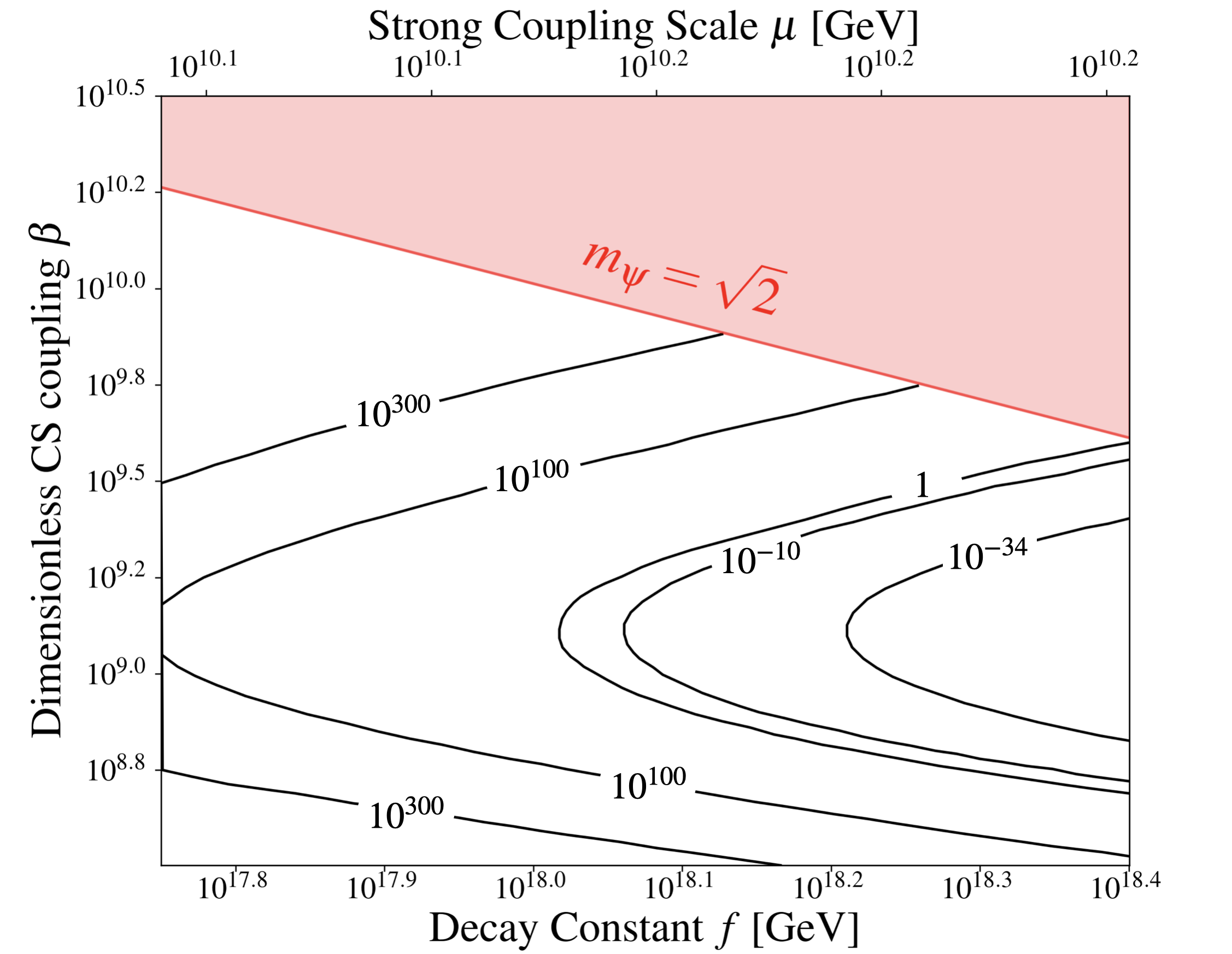}
    \caption{The tensor power spectrum amplitude $\Delta_\gamma^2$ is amplified by the instability for most of the parameter space. The viable region with $\Delta_\gamma^2 < \mathcal{O}(10^{-7})$ is near the Planck scale $f \lesssim M_\mathrm{Pl} = 2.4\times 10^{18}~\mathrm{GeV}$. }
    \label{fig:tensor_contour}
\end{figure}

To begin, we identify the parameter space where the amplitudes of tensor perturbation are not hopelessly enhanced by the instability in the left-handed spin-$2$ gauge modes. To ensure the tensor-to-scalar ratio $r \lesssim 0.03$ and the scalar amplitude $A_s \sim 10^{-9}$, the upper limit for the tensor power spectrum amplitude is about $\Delta_\gamma^2\sim\mathcal{O}(10^{-7})$. This confines the viable parameter space to a small region where $m_\psi$ and $m_v$ are of order $\mathcal{O}(1 \sim 10)$. In the crude estimation of $m_\psi$ and $m_v$ in Sec.~\ref{sec:bacground}, the viable parameter has the decay constant $f \gtrsim 10^{18}~\mathrm{GeV}$, CS coupling $\beta \sim 10^9$, and the strong coupling scale $\mu \gtrsim 10^{10}~\mathrm{GeV}$. This agrees with our numerical results in figure~\ref{fig:tensor_contour}. It is worth mentioning that this area is many orders of magnitude away from the electric dipole moment and $N=60$ constraints.  Attempts of relaxing these two constraints would not help to broaden the viable parameter space.
\\
\noindent
We thus focus on the region with small tensor power spectrum amplitude $\Delta_\gamma^2$ and ignore the rest of the parameter space. We sample a grid within the $\mathcal{O}(10^{-8})$ contour line of the tensor amplitude. For each point on the parameter space, we evolve the background fields to determine the pNG Higgs angle at horizon crossing, \textit{i.e.,} $60$ $e$-foldings prior to the end of inflation. We then fix the Hubble parameter $H$, the background gauge and Higgs fields values $\psi$ and $\bar{h}$, and the two effective gauge masses $m_\psi$ and $m_v$. Given all the background parameters, we evolve the scalar perturbations $\Delta = (\h, \hat{\varphi}, \hat{z},\hat{X})$ numerically using their equations of motion in Appendix~\ref{sec:scalareom}. The initial time is set to be $x_i = 100\times \mathbf{\Lambda}\bar{h}m_\psi$, which ensures the WKB approximations of the initial conditions in Eqs.~\ref{eq:wkb_initial} are valid, and the final time corresponds to $x_f=10^{-3}$. We extrapolate the late-time behaviors of the scalar perturbations using power laws and evaluate the scalar and tensor power spectrum amplitudes at $x = 0.02$.

\begin{figure}[h]
    \centering
    \includegraphics[width=0.9\linewidth]{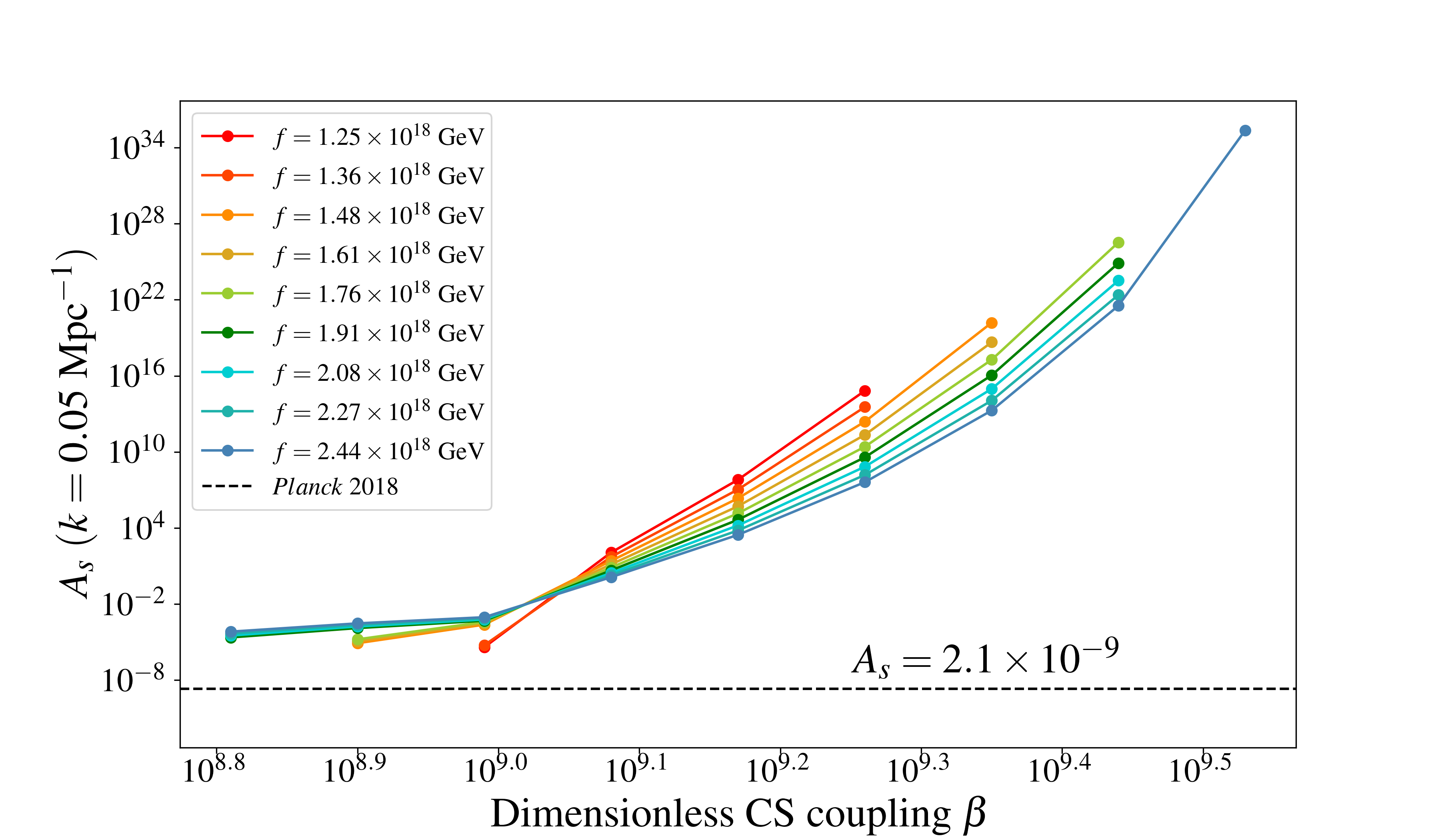}
    \caption{Scalar power spectrum amplitude $A_s$ within the $\mathcal{O}(10^{-8})$ contour of tensor power spectrum amplitude $A_t$. We evaluate the amplitudes at $k = 0.05~\mathrm{Mpc}$ corresponding to $60$ $e$-folds before the end of the inflation. The reddish lines represent lower decay constants, while the blueish lines represent higher decay constants. The points are sampled in the area with tensor power spectrum amplitude $A_t < \mathcal{O}(10^{-8})$, \textit{i.e.,} the area without the overproduction of tensor perturbations. The numerical evaluations throughout the area show that the scalar power spectrum amplitude $A_s$ decreases as the Chern-Simons coupling $\beta$ decreases. However, even with the optimal choices of parameters ($f = 1.25 \times 10^{18}~\mathrm{GeV}$ and $\beta = 10^9$), the scalar power spectrum amplitude is still three orders of magnitudes larger than the Planck measurement.}
    \label{fig:As}
\end{figure}

The result of our numerical analysis is presented in figure~\ref{fig:As}. We find that the scalar perturbations are inevitably overproduced within the $\mathcal{O}(10^{-8})$ contour line of the tensor amplitudes. To reduce the scalar perturbations, lower values of the decay constant $f$ and Chern-Simons coupling $\beta$ are needed. However, they would increase the effective gauge masses $m_\psi$ and $m_v$ such that exponentially enhances the tensor power spectrum amplitudes. We find that the minimal pNG Higgs inflation based on the $\mathrm{SU(5)/SO(5)}$ symmetry coset either overproduces scalar perturbations or tensor perturbations or both at the same time. 

\section{Discussion and Conclusion}
The instability in the left-handed spin-$2$ gauge modes is a generic feature of the chromo-natural inflation \cite{Adshead:2012kp,Adshead:2013nka,Dimastrogiovanni:2012ew} and many of its variants (e.g. \cite{Adshead:2016omu, Obata:2014loa}). The exponential growth of the tensor perturbations induced by the instability leads to a surge in the tensor-to-scalar ratio $r$, and any attempt to lower the tensor-to-scalar ratio red-tilts the spectrum too severely. The predicament of retaining a healthy spectrum and lowering the tensor-to-scalar ratio $r$ in the end rules out the original version of the chromo-natural inflation. One can suppress the tensor-to-scalar ratio by spontaneously breaking the $\mathrm{SU(2)}$ gauge symmetry such that a new Higgs degree of freedom amplifies the scalar perturbations. Although the Higgs mode also amplifies the tensor modes, the increase is much smaller in comparison with the scalar perturbations. The Higgsed CNI is shown to be able to align with CMB measurements. 

In an EFT perspective, our pNG Higgs inflation is similar to the Higgsed chromo-natural inflation except for the Chern-Simons term
\begin{align}
    \mathcal{L}_\mathrm{CS} = \frac{\lambda}{f}\phi F_{\mu\nu}^a \Tilde{F}^{a\mu\nu} \longrightarrow \left(\frac{g\beta}{f}\right)^2 h^2 F_{\mu\nu}^a \Tilde{F}^{a\mu\nu},
\end{align}
where $\lambda$ is the Chern-Simons coupling for CNI and $\phi$ here is the axion. We stress that there are two crucial differences: the $\mathrm{SU(2)}$ gauge in the pNG Higgs inflation is the Standard Model weak isospin gauge boson, while the $\mathrm{SU(2)}$ gauge in CNI is exotic; the dimension-six operator in the pNG Higgs inflation is suppressed by $f^2$, but the dimension-five operator in CNI is only suppressed by $f$.

The goal of connecting cosmic inflation and electroweak symmetry break introduces a touch of elegance to the Higgs inflation, but it also confines the model by fixing the Higgs VEV at $v_\mathrm{EW} \approx 246~\mathrm{GeV}$. The tensor instability is only tuned by the effective gauge masses $m_\psi$ and $m_v$ and the only free parameter in $m_v$, as shown in Eq.~(\ref{eq:m_approx}), is the strong coupling scale $\mu$. Thus, to have $m_v \sim \mathcal{O}(10)$ fixes the strong coupling scale at $\mu\sim 10^{10}~\mathrm{GeV}$ and also the decay constant at $f\sim 10^{18}~\mathrm{GeV}$ so that the model predicts the correct Higgs mass $m_H \approx 125~\mathrm{GeV}$. Furthermore, the quadratic Higgs-Chern-Simons term elevates the impacts of large coupling constants. Eq.~(\ref{eq:mpsi-approx}) shows that the approximated $m_\psi$ is linearly proportional to the decay constant $f$ in a theory with dimension-six operator, while a dimension five operator would mitigate $m_\psi$ by a factor of $\sqrt{f}$. However, there is no gauge-invariant way of writing down a dimension five Higgs-Chern-Simons operator.

This work studies the minimal set-up of the pNG Higgs inflation and paves the ways for its variants. We remind the readers that the construction of the pNG Higgs, by itself, presents a rich ensemble of scalar degrees of freedom \cite{Alexander:2023flr}. In the `pion' matrix of $\mathrm{SU(5)/SO(5)}$ littlest Higgs
\begin{align}
\Pi=\begin{pmatrix}
\omega-\frac{\eta}{\sqrt{20}} \mathbb{I}_2 & \frac{H}{\sqrt{2}} & \phi^{\dagger} \\
\frac{H^{\dagger}}{\sqrt{2}} & \sqrt{\frac{4}{5}} \eta & \frac{H^{\top}}{\sqrt{2}} \\
\phi & \frac{H^*}{\sqrt{2}} & \omega^{\top}-\frac{\eta}{\sqrt{20}} \mathbb{I}_2
\end{pmatrix}
\end{align}
we recognize the complex doublet $H$ as the Higgs, while there are $10$ more degrees of freedom: a real triplet $\omega$, a real scalar $\eta$, and a complex triplet $\phi$. We explicitly break the $SU(5)$ by gauging a $\mathrm{SU(2)\times[SU(2)\times U(1)]}$ subgroup. Subsequently, this residual symmetry is spontaneously broken into the Standard Model $\mathrm{SU(2)_L \times U(1)_Y}$ such that the two $\mathrm{SU(2)}$ gauges eat the real triplet $\eta$ and scalar $\omega$, yielding three massive bosons, three massless bosons, and a massless Goldstone mode. In our previous analysis, we neglected the Goldstone mode and the complex triplet $\phi$. 

These inherent scalar fields in the pNGB Higgs coset make many interesting multi-field dynamics possible and thus resolve the tension with the CMB measurements. An intuitive extension of our pNG Higgs inflation would be a two-pNGB inflation similar to \cite{Obata:2014loa}. In this scenario, inflation is initiated by a non-Higgs pseudo scalar and the effects of gauge field are negligible in this phase. Since the overproduction of tensor perturbations is essentially sourced by the left-handed spin-$2$ gauge mode, an effectively inactive gauge sector can potentially suppress the effective gauge mass $m_\psi$ and relieve the model's tension with CMB observations. After tens of $e$-folds, the pNG Higgs takes over driving inflation, and the Higgs-gauge dynamics can produce chiral gravitational waves under the mechanism discussed in Sec.~\ref{sec:tensor}. Another well-studied extension of CNI is the spectator field CNI (S-CNI) model \cite{Dimastrogiovanni:2016fuu}, which the pNG Higgs is no longer the slow-roll inflaton but rather part of a spectator sector. As another scalar field is in charge of inflation, the slow-rolling of the pNG Higgs is no longer mandatory such that the viable parameter space can be much wider and include a small decay constant $f$ and small Chern-Simons coupling $\beta$. Indeed, the chiral gravitational wave sourced by the tensor instability can imprint on the CMB. We leave the details of model-building and phenomenological investigations for future works.

\section*{Acknowledgment}
We thank Peter Adshead, Tom Giblin, and Geoffrey Beck for valuable discussions. SA and CN was supported by the Simons Foundation award number 896696. CN was partially supported by Brown physics department. HBG is supported by the National Science Foundation, MPS Ascending Fellowship, Award 2213126.

\appendix
\section{Full Details of Scalar Perturbations}\label{sec:scalareom}

The full equation of motion for the scalar perturbations $\Delta = (\h, \hat{\varphi}, \hat{z},\hat{X})$ are given by
\begin{align}
    &\begin{aligned}
        \h'' + & \left(1-\frac{2}{x^2}+\frac{V_{, h h}}{x^2 H^2}+ \frac{\Lambda^2\bar{h}^2 m_\psi^2}{2 m_\psi^2 + m_v^2+x^2}-\frac{3 \Lambda m_\psi \psi}{x^2}\right) \h   \\
        & + \frac{2 \sqrt{2} \Lambda m_\psi \bar{h}}{x^2} \hat{z} + \frac{\sqrt{2} \Lambda m_\psi \bar{h}}{x} \hat{z}' -\frac{\sqrt{2}\Lambda m_\psi \Bar{h}}{x} \left(\frac{x^2}{m_\psi^2}+2\right)^{-1 / 2} \hat{\varphi}'  \\
        &-\frac{\sqrt{2}\Lambda \bar{h} \left(8 m_\psi^6+6 m_\psi^4 x^2+3 m_\psi^2 x^4+x^6+m_v^2 m_\psi^2(4 m_\psi^2+x^2)\right)}{m_\psi x^2(2 m_\psi^2+x^2)(m_v^2+2 m_\psi^2+x^2)}\left(\frac{x^2}{m_\psi^2}+2\right)^{-1 / 2}\hat{\varphi} \\
        & + \frac{\Lambda \Bar{h}m_v m_\psi}{\sqrt{(2m_\psi^2+x^2)(m_v^2+2m_\psi^2+x^2)}} \hat{X}' + \frac{\Lambda \Bar{h}m_v m_\psi\left((2 m_\psi^2+x^2)^2+2 m_v^2 m_\psi^2\right)}{x(m_v^2+2 m_\psi^2+x^2)^{3/2}(2 m_\psi^2+x^2)^{3/2}}\hat{X} = 0
    \end{aligned}\\
    &\begin{aligned}
       &\hat{\varphi}'' + \left(1-\frac{2}{2m_\psi^2+x^2} + \frac{2m_\psi^2}{x^2} + \frac{m_\psi^2m_v^2}{2x^2(m_\psi^2 + x^2)}+\frac{2m_\psi^2\left(3-\frac{4m_v^2}{m_v^2+2m_\psi^2+x^2}\right)}{(m_\psi^2 + x^2)^2}\right)\hat{\varphi} \\
        &+ \frac{(2+m_v^2)\sqrt{2+\frac{x^2}{m_\psi^2}}}{x^2}\hat{z} + \frac{\sqrt{2}\mathbf{\Lambda} m_\psi \Bar{h}}{x} \left(\frac{x^2}{m_\psi^2}+2\right)^{-1 / 2} \h' - \frac{\Lambda \Bar{h}\left(3 m_v^2 m_\psi^2+6 m_\psi^4+3 m_\psi^2 x^2+x^4\right)}{m_\psi x^2\left(m_v^2+2 m_\psi^2+x^2\right) \sqrt{1+\frac{x^2}{2 m_\psi^2}}}\h \\
     & + \frac{\sqrt{2} m_vm_\psi\left(m_v^4(2m_\psi^2+x^2) + (2m_\psi^2+x^2)^2(2+2m_\psi^2+x^2)+2m_v^2\left[4m_\psi^4 + x^4 + m_\psi^2(2+4x^2)\right]\right)}{x(2m_\psi^2+x^2)^2(m_v^2+2m_\psi^2+x^2)^{3/2}} \hat{X} \\
    & - \frac{2\sqrt{2}m_v m_\psi}{(2m_\psi^2+x^2)\sqrt{m_v^2+2m_\psi^2+x^2}} \hat{X}'
    \end{aligned} 
\end{align}
\begin{align}
    & \hat{z}'' + \frac{3\sqrt{2} \mathbf{\Lambda} m_\psi \bar{h}}{x^2}\h  - \frac{\sqrt{2} \mathbf{\Lambda} m_\psi \bar{h}}{x}\h' + \left(1-\frac{2-2 m_\psi^2}{x^2}\right) \hat{z} + \frac{(2+m_v^2) \sqrt{2+\frac{x^2}{m_\psi^2}}}{x^2} \hat{\varphi} =0\\
    &\begin{aligned}
        \hat{X}'' &+ \left(1 - \frac{2}{x^2} + \frac{m_v^2(1+2m_\psi^2 + m_v^2+x^2)}{(2m_\psi^2 + x^2)(m_v^2 + 2m_\psi^2 + x^2)} + \frac{m_v^2x^2(m_v^2+8m_\psi^2+4x^2)}{(2m_\psi^2 + x^2)^2(m_v^2 + 2m_\psi^2 + x^2)^2}\right)\hat{X}  \\
        &- \frac{2\sqrt{2}m_v m_\psi}{(2m_\psi^2+x^2)\sqrt{m_v^2+2m_\psi^2+x^2}} \hat{\varphi}' - \frac{\Lambda \Bar{h}m_v m_\psi}{\sqrt{(2m_\psi^2+x^2)(m_v^2+2m_\psi^2+x^2)}}\h' \\
        & - \frac{\sqrt{2} m_vm_\psi\left(4m_\psi^4 + x^2(x^2+m_v^2-4)+2m_\psi^2(2x^2+m_v^2+2)\right)}{x(2m_\psi^2+x^2)^2(m_v^2+2m_\psi^2+x^2)^{1/2}}\hat{\varphi} \\
        & + \frac{\Lambda \Bar{h}m_\psi m_v\left(m_v^2+2m_\psi^2+3x^2\right)}{x\left(m_v^2+2m_\psi^2+x^2\right)^{3/2}\left(2 m_\psi^2+x^2\right)^{1/2}}\h = 0,
    \end{aligned}
\end{align}
where the parameters $m_\psi, m_v, \mathbf{\Lambda}$, and background fields $\bar{h}$ and $\psi$ are fixed at their horizon-crossing values for the numerical evaluations. This approximation is valid as the background dynamic is negligibly small for the physics of interest.

\bibliographystyle{JHEP}
\bibliography{perturbation.bib}

\end{document}